\documentclass[%
 reprint,
 amsmath,amssymb,
 aps,
]{revtex4-1}

\usepackage{graphicx}
\usepackage{dcolumn}
\usepackage{bm}
\usepackage{bbold}
\usepackage{xspace}
\usepackage{mathtools}
\usepackage{tikz}
\usepackage{xcolor}
\usepackage{lipsum}
\usepackage{soul}

\usetikzlibrary{automata,arrows,positioning,calc}
\newcommand{\be}{\begin{equation}}
\newcommand{\ee}{\end{equation}}
\newcommand{\ba}{\begin{eqnarray}}
\newcommand{\ea}{\end{eqnarray}}
\newcommand{\bi}[1]{Fig.~\ref{fig:#1}}

\newcommand{\lr}[1]{\left\langle #1 \right\rangle}

\newcommand{\fig}[1]{Fig.~\ref{#1}}

\newcommand{\eq}[1]{eq.~(\ref{#1})}
\newcommand{\eqs}[1]{eqs.~(\ref{#1})}
\newcommand{\tab}[1]{Table~\ref{#1}}
\newcommand{\Na}[0]{Na\textsuperscript{+}\xspace}
\newcommand{\Ka}[0]{K\textsuperscript{+}\xspace}
\newcommand{\Ca}[0]{\text{Ca}\textsuperscript{2+}\xspace}
\newcommand{\Ip}[0]{\text{IP}\textsubscript{3}\xspace}
\newcommand{\mean}[1]{\langle #1 \rangle}
\newcommand{\se}[1]{sec.~(\ref{#1})}

\begin{document}

\title{
The noise intensity of a Markov chain}

\author{Lukas Ramlow}
\author{Benjamin Lindner} 
\affiliation{Bernstein Center for Computational Neuroscience Berlin, Philippstr.\ 13, Haus 2, 10115 Berlin, Germany}
\affiliation{Physics Department of Humboldt University Berlin, Newtonstr.\ 15, 12489 Berlin, Germany}
\date{\today}

\begin{abstract}
Stochastic transitions between discrete microscopic states play an important role in many physical and biological systems. Often, these transitions lead to fluctuations on a macroscopic scale.
A classic example from neuroscience is the stochastic opening and closing of ion channels and the resulting fluctuations in membrane current. When the microscopic transitions are fast, the macroscopic fluctuations are nearly uncorrelated and can be fully characterized by their mean and noise intensity. We show how, for an arbitrary Markov chain, the noise intensity can be determined from an algebraic equation, based on the transition rate matrix. 
We demonstrate the validity of the theory using an analytically tractable two-state Markovian dichotomous noise, an eight-state model for a Calcium channel subunit (De Young-Keizer model), and Markov models of the voltage-gated Sodium and Potassium channels as they appear in a stochastic version of the Hodgkin-Huxley model.
\end{abstract}

\maketitle


\section{\label{sec:level1}Introduction}
For models of many fluctuation phenomena in physics, biology, chemistry, and other fields, it is important to properly characterize the noise that drives a given dynamical system. 
Examples include the firing of a neuron, driven by channel noise \cite{WhiRub00, DorWhi05, SchFis10, FisSch12, MoeIan16, PuTho21} and by shot-noise-like input from other neurons \cite{HohBur01, RicGer06, WolLin10, RicSwa10, BriDes15, DroLin17}, the fluctuations in the intensity of an excitable laser \cite{HakSau67, DubKra99, LamGuz03, SchTia12}, or chemical reactions in mesoscopically small volumes \cite{Gar85, Van92, Gil00}. 
In the past decades, strongly simplified noise models, such as white Gaussian noise, Poissonian shot noise, dichotomous noise, or an exponentially correlated Ornstein-Uhlenbeck noise have often been used to describe the input noise in these systems. 
As the models for the driving noise process become more complex, one would like to use characteristics of the noise process that can be used to fairly compare different noise models (and their effect on a dynamical system). This fair comparison is already possible for simple noise processes, such as different exponentially correlated processes like the Gaussian Ornstein-Uhlenbeck process \cite{UhlOrn30, Ris84,HanJun95}, the dichotomous telegraph noise \cite{HorLef83, Gar85, Ben06,DroLin14}, or the exponentially distributed noise \cite{FarLin21} (which is, however, only approximately exponentially correlated).  

Simple characteristics of a stochastic (noise) process $x(t)$ are its stationary mean and variance 
\begin{align}
\mu = \lr{x(t)},\;\;\; \sigma^2=\lr{\Delta x^2}=\lr{x^2(t)}-\lr{x(t)}^2,
\label{eq:mean_variance}
\end{align}
its correlation time
\begin{align}
\tau =\int_0^\infty dt' \frac{\lr{x(t)x(t+t')}-\lr{x(t)}^2}{\lr{\Delta x^2}},
\label{eq:correlation-time}
\end{align}
and its noise intensity
\begin{align}
D =\int_0^\infty dt'\lr{x(t)x(t+t')}-\lr{x(t)}^2.
\label{eq:noise-intensity}
\end{align}
The meaning of mean and variance are quite obvious. The correlation time (here defined by an integral over the normalized autocorrelation function) provide an order-of-magnitude estimate of the periods over which the process changes significantly.
Last but not least, the intensity (here defined by an integral over the \emph{un}normalized autocorrelation function) captures how much of an effect the process would have when driving a dynamical system. More specifically, if $x(t)$ were the velocity of a Brownian particle, $D$ would correspond to the diffusion coefficient, which is a reasonable measure of the effect of the velocity noise on the position dynamics. 

It is clear from the above definitions that correlation time, variance, and intensity are connected by
\begin{align}
D =\sigma^2 \tau,
\label{eq:intensity_variance_correlation-time}
\end{align}
i.e.\ if we know two of the characteristics, we can easily compute the third one. 
For processes described by a nonlinear Langevin equation, all four characteristics (including also the mean value) can be expressed by quadratures \cite{Ris89}; see also refs.~\cite{Lin07, Lin08}, which make the above-mentioned connection between noise intensity and diffusion coefficient more explicit.

For discrete-valued processes, $x(t) \in \{x_1,x_2,\dots \}$, governed by a master equation, the mean and the variance can easily be calculated in all cases where the stationary probability can be obtained. 
The calculation of the noise intensity is more involved but has recently been worked out in our previous paper and applied for a specific model \cite{RamFal23,RamFal23b}. 
It turned out that, in complete analogy to the procedure for the Langevin case studied in \cite{Ris89, Lin07}, we can obtain closed-form expressions for these characteristics. Multiple integrals in the continuous Langevin case correspond to multiple sums in the Markov chain case. Here, we show that the theory is not limited to a specific model, but can be applied to any random process where the state transitions are governed by a master equation.  

Our paper is structured as follows. We begin in \se{sec:noise-intensity} with the general framework for calculating the noise intensity. Then we discuss three examples. In \se{sec:dichotomous_noise} we illustrate the general result for the (simple and well-known) case of Markovian dichotomous noise. In \se{sec:de-young-keizer} we study the more complicated case of an eight-state Markov model, as used in the De Young-Keizer model to describe a subunit of calcium release channel. In \se{sec:potassium-sodium-channels} we study a stochastic version of
the sodium and potassium currents as they appear in the Hodgkin-Huxley model with channel noise.
Finally, in \se{sec:discussion}, we discuss further applications of our results, such as a general white-noise approximation in cases where the microscopic transitions are much faster than the dynamics of the driven system. However, we also point out some limitations for systems where the noise is \emph{not} purely external and/or very fast but also depends on the state of the driven system.

\section{Noise intensity of a Markov chain}
\label{sec:noise-intensity}
We consider a random process $x(t)$ with discrete states, where the probability $p_i(t)$ of finding a state $i$ at time $t$ is determined by the (homogeneous) master equation:
\begin{align}
    \dot{\boldsymbol{p}}(t) = W \boldsymbol{p}(t)
    \label{eq:master_equation}
\end{align}
with the probability vector $\boldsymbol{p}(t) = \begin{pmatrix} p_1(t) & p_2(t) & \dots \end{pmatrix}$ and the transition rate matrix $W = (w_{ij})$. The entries $w_{ij} > 0$ for $i\neq j$ are the transition rates from a state $j$ to state a $i$ and $w_{jj} = - \sum_{i \neq j} w_{ij}$. To fully characterize the process $x(t)$, each state $i$ is assigned a specific value $x_i$, which are not necessarily different from each other.

For such a random process, the calculation of the mean and the variance follows standard procedures \cite{Gar85, Van92} and is based on the stationary probability vector $\lim_{t\to\infty}\boldsymbol{p}(t) = \boldsymbol{p}$ (we indicate the stationary state by omitting the time argument). This vector can be obtained from the stationary master equation
\begin{align}
0 = W \boldsymbol{p}
\label{eq:stationary_master_equation}
\end{align}
together with the normalization condition $\sum_i p_i = 1$; the additional condition is needed because of the rank deficiency of the matrix $W$. Practically, the normalization can be incorporated by replacing an arbitrary row of the matrix $W$ with ones and the corresponding entry in the zero vector on the l.h.s.\ by a one. This leads to a linear system of equations solvable by standard methods. 
Given the stationary probabilities $p_i$, the mean and the variance can be calculated by
\begin{align}
\mu &= \sum x_i p_i, & \sigma^2 &= \sum_i (x_i - \mu)^2 p_i.
\label{eq:mean_and_variance}
\end{align}

The calculation of the noise intensity is more advanced. We recall the definition of the noise intensity by the integral over the autocorrelation function \eq{eq:noise-intensity}. 
In principle, the correlation function can be determined by solving the time-dependent master \eq{eq:master_equation}. However, it turns out that the calculation of the time-dependent probability vector $\boldsymbol{p}(t)$ is not necessary in order to calculate the noise intensity. Instead, taking advantage of the fact that the integrated correlation function is of interest, an algebraic equation can be found that determines the noise intensity and is not much more complicated to solve than the equations that determine the mean or variance.

To show this, we relate the noise intensity to the probabilities of the Markov chain:
\begin{equation}
	\begin{aligned}
		D &= \int_{0}^{\infty} dt' \, \mean{x(t+t')x(t)} - \mean{x(t)}^2 \\
        &= \int_{0}^{\infty} dt'\, \sum_{i, j}  [x_i x_j p_{ij}(t')p_j - 
         x_i x_j p_i p_j] \\
		&= \sum_{i,j} x_i f_{ij} x_j p_j.
		\label{eq:noise-intensity_sum}
	\end{aligned}
\end{equation}
where $p_{ij}(t') = p(i, t+t' | j, t)$ is the transition probability, i.e.\ the probability of finding the state $i$ at $t+t'$ given the state $j$ at time $t$. Since we are considering a homogeneous process, this conditional probability does not depend on the absolute time $t$, but only on the difference $t'$ and is determined by the master \eq{eq:master_equation} with the initial condition $p_k(t) = \delta_{kj}$. The auxiliary function introduced in the last line of \eq{eq:noise-intensity_sum},
\begin{align}
    f_{ij} = \int_{0}^{\infty} dt'\, p_{ij}(t') - p_i,
    \label{eq:auxiliary_function}
\end{align}
is given by the integral over the difference between the transition and stationary probabilities. Eq.\ (\ref{eq:noise-intensity_sum}) is of course just a reformulation of the problem. However, it turns out that the auxiliary functions $f_{ij}$ for a given $j$ can be calculated from a system of algebraic equations together with an additional condition, a calculation that is very similar to that of the stationary probabilities $p_i$. 
To see this, we formulate the master equation where the state at some reference time $t$ has been specified ($p_{kj}(0) = \delta_{kj}$):
\begin{equation}
\begin{aligned}
		\dot{p}_{kj}(t') &= \sum_{i} w_{ki}p_{ij}(t'), \\
        \dot{p}_{kj}(t') &= \sum_{i} w_{ki}[p_{ij}(t') - p_i], \\
		p_k - \delta_{kj} &= \sum_{i} w_{ki} f_{ij}.
	\end{aligned}
         \label{eq:auxiliary_function_sum}
\end{equation}
To get from the first to the second line we have subtracted the stationary master equation $0 = \textstyle\sum_{i} w_{ki} p_i$. To get from the second to the third line we have integrated over $t'$, used \eq{eq:auxiliary_function_sum}, and exploited that $\int_0^\infty dt'\, \dot{p}_{kj}(t') = p_k - \delta_{kj}$. The last line in \eq{eq:auxiliary_function_sum} looks like an equation that uniquely determines $f_{ij}$. However, because of the rank deficiency of $W$ we need an additional condition that is obtained by observing that
\begin{align}
    \sum_i f_{ij} = \int_0^\infty dt' \, \sum_i p_{ij}(t') - p_i = 0.
    \label{eq:auxilliary_condition}
\end{align}
This condition is independent of $j$ and reflects that for any $t'$ both the transition probability and the stationary probability are normalized over the states $i$. 

Finally, while \eqs{eq:noise-intensity_sum} - (\ref{eq:auxilliary_condition}) allow for the calculation of noise intensity, they can be expressed more conveniently. For this purpose, we write \eq{eq:noise-intensity_sum} in matrix notation
\begin{align}
    D = \boldsymbol{x}^T  F \boldsymbol{y},
    \label{eq:noise-intensity_matrix}
\end{align}
where $f_{ij}$ is the entry in the $i$-th row and $j$-th column of the matrix $F$ and the two vectors are given by $\boldsymbol{x} = \begin{pmatrix} x_1 & x_2 & ... \end{pmatrix}^T$ and $\boldsymbol{y} = \begin{pmatrix} x_1 p_1 & x_2 p_2 & ... \end{pmatrix}^T$.
Similarly, the set of linear \eqs{eq:auxiliary_function_sum} can be combined into a single matrix equation
\begin{align}
    P - \mathbb{1} = W F
    \label{eq:master_equation_F}
\end{align}
where $P$ is a matrix in which each entry in the $i$-th row is given by the stationary probability $p_i$ and $\mathbb{1}$ is the identity matrix. The additional condition \eq{eq:auxilliary_condition} can be written as 
\begin{align}
    \begin{pmatrix}
        1 & 1 & \dots & 1
    \end{pmatrix} F = 0,
\end{align}
implying that each column of the matrix $F$ adds up to zero. Again, practically, these conditions can be incorporated by replacing an arbitrary row in $W$ by ones and the corresponding row of the matrix $P - \mathbb{1}$ by zeroes.

In the following, we put the theory to the test for different models. The first model is the well-known Markovian two-state model, the second is a more involved eight-state model for a subunit of a Calcium channel (De Young-Keizer model), the third example comprises the Sodium and Potassium channels in a stochastic version of the Hodgkin-Huxley model of an excitable nerve membrane.

\section{A simple example:\\ Markovian dichotomous noise}
\label{sec:dichotomous_noise}
\begin{figure}
    \centering
    \includegraphics[scale=1.25]{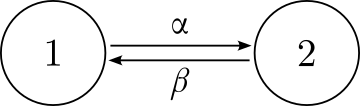}
    \caption{\textbf{State diagram of a dichotomous noise.} The system consists of two states $1$ and $2$ with corresponding levels $x_1$ and $x_2$. The transitions between the two states occur at rates $\alpha$ and $\beta$.}
    \label{fig:1-diagram_dichotomous}
\end{figure}

As an introduction to the method, we consider a Markovian dichotomous noise for which the noise intensity is known and can be calculated in several ways \cite{HorLef83}. A dichotomous Markov noise $x(t)$ is a Markov process with two levels $x_1$ and $x_2$, corresponding to two different states with transition rates $\alpha$ and $\beta$ between them. A schematic representation of the model is shown in \fig{fig:1-diagram_dichotomous}.

To calculate the noise intensity according to \eq{eq:noise-intensity_matrix}, we first determine the stationary probabilities using the stationary master \eq{eq:stationary_master_equation}
\begin{align}
0  = \begin{pmatrix}
   -\alpha & \beta\\
   \alpha & -\beta
\end{pmatrix}
\begin{pmatrix}
    p_1 \\
    p_2
\end{pmatrix},
\end{align}
together with the normalization condition $p_1 + p_2 = 1$ and obtain $p_1 = \beta / (\alpha + \beta)$ and $p_2 = \alpha / (\alpha + \beta)$.
We can now calculate $F$ using \eq{eq:master_equation_F}
\begin{equation}
\begin{aligned}
    \begin{pmatrix}
        p_1 & p_1 \\
        p_2 & p_2 \\
    \end{pmatrix}
    &- 
    \begin{pmatrix}
    1 & 0 \\
    0 & 1
    \end{pmatrix} 
    = \\
    &
    \begin{pmatrix}
        -\alpha & \beta\\
        \alpha & -\beta
    \end{pmatrix}
    \begin{pmatrix}
        f_{11} & f_{12} \\
        f_{21} & f_{22} \\
    \end{pmatrix}
\end{aligned}
\end{equation}
with the additional conditions that each column of $F$ sums to zero, i.e.\ $f_{11} + f_{21} = 0$ and $f_{12} + f_{22} = 0$.
This yields:
\begin{align}
    F = \frac{1}{(\alpha + \beta)^2}\begin{pmatrix}
        \alpha & - \beta \\
        -\alpha & \beta
    \end{pmatrix}
\end{align}
and allows to determine the noise intensity using \eq{eq:noise-intensity_matrix}:
\begin{equation}
\begin{aligned}
    D &= 
    \frac{1}{(\alpha + \beta)^3}
    \begin{pmatrix}
        x_1 & x_2 
    \end{pmatrix}
    \begin{pmatrix}
        \alpha & - \beta \\
        -\alpha & \beta
    \end{pmatrix}
    \begin{pmatrix}
        x_1 \beta \\
        x_2 \alpha \\
    \end{pmatrix} 
    \\
    &= \frac{\alpha \beta}{(\alpha + \beta)^3}  (x_1 - x_2)^2
\end{aligned}
\end{equation}
an expression that is in agreement with the result presented in \cite{HorLef83}. 
\begin{figure}
    \centering
    \includegraphics[width=\columnwidth]{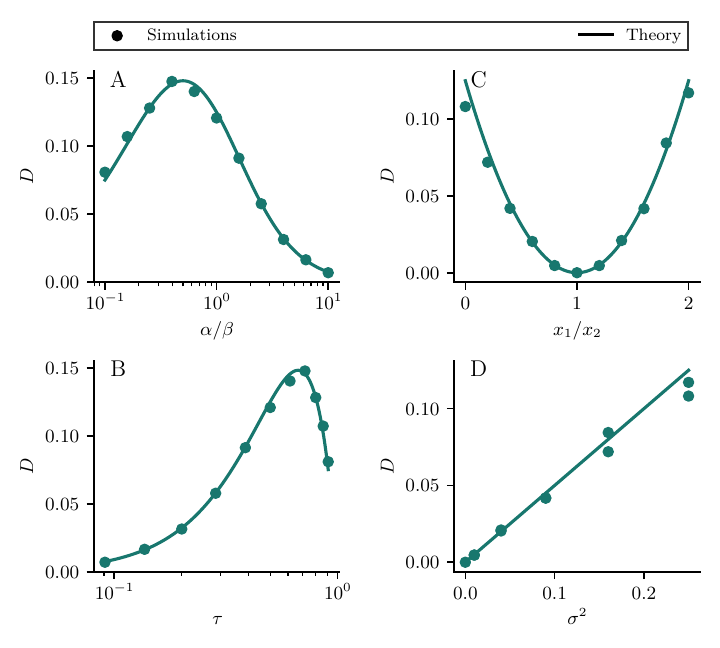}
    \caption{\textbf{The noise intensity of a Markovian Dichotomous noise.} Panels A and B show the noise intensity as a function of the ratio between the transition rates $\alpha / \beta$ or the correlation time $\tau = 1/(\alpha + \beta)$. The intensity has a maximum at $\alpha^* = \beta / 2$ or $\tau^* = 2/(3\beta)$. Panels C and D show the noise intensity as a function of the ratio between the values taken in the two states $x_1 / x_2$ or the variance $\sigma^2 = \alpha \beta (x_1 - x_2)^2 / (\alpha + \beta)^2$. The intensity has a minimum at $x_1^* = x_2$ and scales linearly with the variance. Parameters: $\beta = 1$, $x_2 = 1$.}
    \label{fig:2-intensity_dichtomous}
\end{figure}
The noise intensity has a maximum as a function of the rate $\alpha$, keeping the other rate $\beta$ fixed, at $\alpha^* = \beta /2$ (\fig{fig:2-intensity_dichtomous}A). As a function of the correlation time $\tau = 1/(\alpha + \beta)$ has also a maximum at $\tau^* = 2/(3 \beta)$ (\fig{fig:2-intensity_dichtomous}B). 
If the intensity is plotted as a function of the ratio $x_1 / x_2$ (\fig{fig:2-intensity_dichtomous}C), it has a minimum and vanishes for $x_1 ^* = x_2$ because the variance vanishes in this case. When plotted as a function of the variance $\sigma^2$ (\fig{fig:2-intensity_dichtomous}D), the noise intensity increases linearly according to \eq{eq:intensity_variance_correlation-time}.

The calculation presented here serves only as a sanity check. The real advantage of the method lies in the possibility of calculating the noise intensity for more complicated transition rate matrices, as we will show in the following.

\section{A biophysical example:\\ Stochastic \Ca channel model}
\label{sec:de-young-keizer}
\begin{figure}
    \centering
    \includegraphics[scale=0.9]{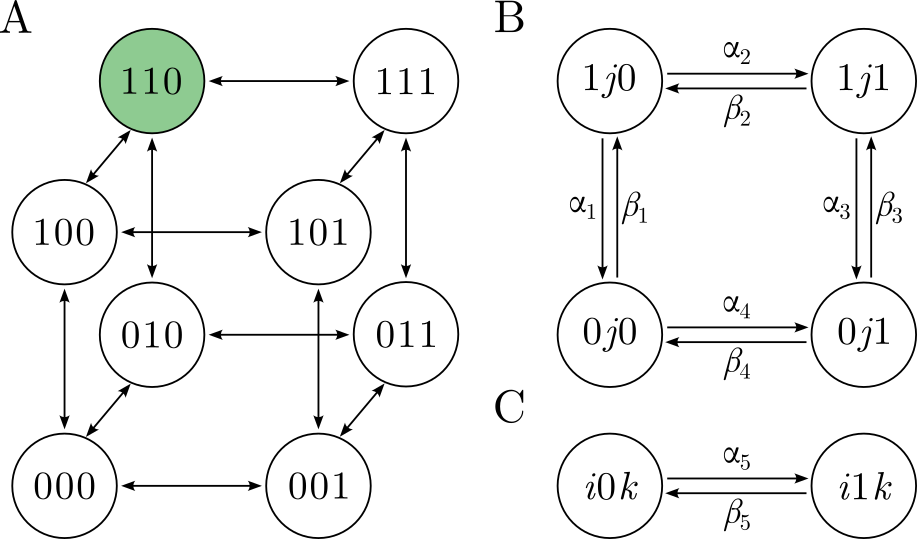}
    \caption{\textbf{State diagram of a \Ca channel subunit \cite{YouKei92}.} Panel A shows the eight-state model of a single \Ip-receptor subunit. The states are denoted $ijk$, where each index represents one of the three binding sites for \Ip ($i$), activating \Ca ($j$), and inhibitory \Ca ($k$). An index is $1$ ($0$) if the binding site is occupied (unoccupied). The conducting state $110$ is highlighted in green. Panel B shows the transition rates on the front and back of the die. Panel C shows the transitions between the front and back faces. Binding rates are denoted $\alpha$ and depend linearly on the corresponding concentration ($\alpha_i = \hat{\alpha}_i[\Ip]$ for $i=1,3$ and $\alpha_i = \hat{\alpha}_i[\Ca]_\text{i}$ for $i=2,4,5$), while unbinding rates are denoted $\beta$ and are constants. Parameters are according to \tab{tab:parameters_deyoung_keizer}.}
    \label{fig:3-diagram-de-young-keizer}
\end{figure}

In this section, we consider a biophysical example and calculate the noise intensity for a eight-state Markov model as illustrated in \fig{fig:3-diagram-de-young-keizer} and used in the De Young-Keizer model to describe a single subunit of an inositol trisphosphate (\Ip) receptor \cite{YouKei92}. For such a model, no closed-form expression for the noise intensity is known.

The entire De Young-Keizer model describes the dynamics of the intracellular calcium (\Ca) concentration, which in many cells serves as a signaling molecule to transmit information about extracellular stimuli (calcium signaling) \cite{BerBoo98, Cla07}. The characteristic short periodic increases in the intracellular \Ca concentration that carry the information can be caused either by an influx of \Ca from the extracellular medium or by a release of \Ca from an intracellular store, the endoplasmic reticulum (ER). In both cases, stochastic transitions between discrete states of the ion channels give rise to macroscopic fluctuations in the intracellular \Ca concentration. 
The De Young-Keizer model covers the case where the \Ca signal is evoked by the release of \Ca from the ER through the \Ip receptor channel.
This receptor channel in turn is assumed to consist of three independent and identical subunits with three binding sites each: one for the second-messenger molecule \Ip, produced in the cell in response to an extracellular stimulus (\Ip pathway), and one activating \Ca binding site and one inhibiting \Ca binding site.
The \Ca current through a single \Ip receptor channel is given by
\begin{align}
    I_\text{Ca} = c_1 x^{(1)}(t)x^{(2)}(t)x^{(3)}(t)([\Ca]_\text{er} - [\Ca]_\text{i}),
\end{align}
where $[\Ca]_\text{er}$ and $[\Ca]_\text{i}$ are the ER and intracellular (cytosolic) \Ca concentrations, $c_1$ is the volume ratio between the ER and the cytosol, and $x^{(n)}(t)$ are dichotomous (two-valued) stochastic processes capturing the state of the three \Ip receptor subunits.

\begin{table}[h]
	\caption{\textbf{Simulation parameters for a \Ip-receptor subunit in the De Young-Keizer model \cite{YouKei92}}}
	\centering
    \begin{tabular}{ l c c }
			\hline
			Parameter & Value & Description \\
			\hline
			\multicolumn{3}{c}{binding constants} \\
			$\hat{\alpha}_1$ / $\mu \text{M}^{-1}\text{s}^{-1}$ & 400 & \Ip  \\
			$\hat{\alpha}_2$ / $\mu \text{M}^{-1}\text{s}^{-1}$ & 0.2 & \Ca inhibition \\
			$\hat{\alpha}_3$ / $\mu \text{M}^{-1}\text{s}^{-1}$ & 400 & \Ip \\
			$\hat{\alpha}_4$ / $\mu \text{M}^{-1}\text{s}^{-1}$ & 0.2 & \Ca inhibition \\
			$\hat{\alpha}_5$ / $\mu \text{M}^{-1}\text{s}^{-1}$ & 20  & \Ca activation \\[0.25cm]
            \multicolumn{3}{c}{dissociation constants $\gamma_i = \beta_i / \hat{\alpha}_i$} \\
            $\gamma_1$ / $\mu \text{M}$ & 0.13 & \Ip \\
            $\gamma_2$ / $\mu \text{M}$ & 1.049 & \Ca inhibition \\
            $\gamma_3$ / $\text{nM}$ & 943.4 & \Ip \\
            $\gamma_4$ / $\text{nM}$ & 144.5 & \Ca inhibition \\
            $\gamma_5$ / $\text{nM}$ & 82.34 & \Ca activation \\
            \hline
		\end{tabular}
	\label{tab:parameters_deyoung_keizer}
\end{table}
The kinetics of a single subunit is described by the scheme shown in \fig{fig:3-diagram-de-young-keizer}. The eight possible states shown in \fig{fig:3-diagram-de-young-keizer}A result from the fact that each of the three binding sites can be in two possible states, occupied or unoccupied. 
The subunit states are labeled $ijk$, where $i$, $j$, and $k$ indicate whether the \Ip, activating \Ca, and inhibitory \Ca binding sites are occupied ($i,j,k = 1$) or unoccupied ($i,j,k=0$). 
The entire channel is open when in all subunits the \Ip and activating \Ca binding sites are occupied and the inhibitory \Ca binding site is unoccupied, i.e.\ when all three subunits are in the $110$ state, highlighted in green in \fig{fig:3-diagram-de-young-keizer}A. 
Put differently, every state is assigned a value according to $x_{ijk} = \delta_{i1}\delta_{j1}\delta_{k0}$, i.e.\ the value of the conducting state $110$ is $1$ and the value of every other state is $0$. 
The transition rates between the states on the front and back faces of the cube are shown in \fig{fig:3-diagram-de-young-keizer}B, while the transition rates between the two faces are shown in \fig{fig:3-diagram-de-young-keizer}C. 
Binding rates are denoted $\alpha$ and depend linearly on the \Ip or \Ca concentration according to the law of mass action ($\alpha_i = \hat{\alpha}_i[\Ip]$ for $i=1,3$ and $\alpha_i = \hat{\alpha}_i[\Ca]_\text{i}$ for $i=2,4,5$), whereas unbinding rates are denoted $\beta$ and are constants (we keep the original notation of De Young and Keizer in terms of $\hat{\alpha}$ and $\gamma = \beta / \hat{\alpha}$, see \tab{tab:parameters_deyoung_keizer}).

While De Young and Keizer considered the \Ip receptor and its subunits in the thermodynamic limit, we calculate the variance $\sigma^2$, correlation time $\tau$ and noise intensity $D$ for a single subunit $x^{(n)}(t)$. 
Although the noise intensity for the Markov chain can be calculated analytically, the expressions for the stationary probability vector $\boldsymbol{p}$ and the auxiliary matrix $F$ are lengthy. Therefore, we determine these two statistics numerically by inverting the transition rate matrix $W$. Since $W$ does not have full rank, this requires some manipulation, that we mentioned already in \se{sec:noise-intensity} and are more detailed below. 

\begin{figure}[t]
    \centering
    \includegraphics[scale=0.8]{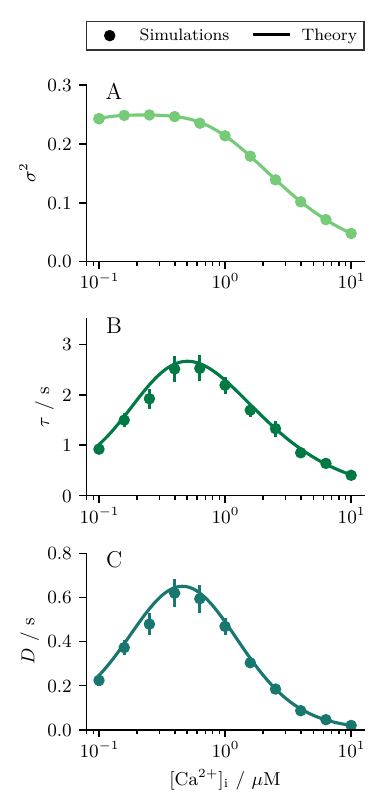}
    \caption{\textbf{Statistical measures of a stochastic \Ca channel subunit.} Panels A, B, and C show the variance $\sigma^2$, correlation time $\tau$, and noise intensity $D$ as a function of the intracellular calcium concenntration $[\Ca]_\text{i}$ for a stochastic \Ip receptor subunit governed by the scheme illustrated in \fig{fig:3-diagram-de-young-keizer}. Vertical lines indicate the standard error calculated from ten simulations. The variance and noise intensity are calculated according to \eq{eq:mean_and_variance} and \eq{eq:noise-intensity_matrix}, respectively. The correlation time is determined as the ratio $\tau = D/\sigma^2$. Parameters are according to \tab{tab:parameters_deyoung_keizer}.}
    \label{fig:4-statistics_de-young-keizer}
\end{figure}
To compute the stationary probability vector $\boldsymbol{p}$, we implement the normalization condition by replacing all entries in an arbitrary row of $W$ with ones, and the corresponding entry in the zero vector on the l.h.s.\ of the stationary master equation (\eq{eq:stationary_master_equation}) with a one. 
This removes a redundant row in $W$, which can be obtained by linear combination of the other rows, and replaces it with the normalization condition $\sum_i p_i = 1$. The same trick is used to compute the auxiliary matrix $F$, again replacing all entries in a row of $W$ with ones and the corresponding row of the matrix $P_0 - \mathbb{1}$ on the l.h.s.\ of \eq{eq:master_equation_F} with zeros. This satisfies the conditions $\sum_i f_{ij} = 0$. 

The results for three statistics are shown in \fig{fig:4-statistics_de-young-keizer}. In all cases, the numerically calculated values show excellent agreement with the theoretical predictions, demonstrating that the method is applicable even when the transition rate matrix is more complicated. Furthermore, the results show that the variance alone is an insufficient measure to quantify the effect of a random process on a driven variable. For example, while the variance is nearly constant for low values of $[\Ca]_\text{i}$, the noise intensity shows a pronounced maximum for an intermediate value.

\section{Voltage-gated channels:\\ Models of stochastic \Ka and \Na channels}
\label{sec:potassium-sodium-channels}
As a third and final example, we calculate the noise intensity for two Markov chains, as used in stochastic variants of the Hodgkin-Huxley model to describe the gating of the potassium (\Ka) and sodium (\Na) channels \cite{FoxLu94, Fox97, Koc99}. Similar to the example discussed in the previous section, microscopic transitions between different discrete states of the ion channels lead to stochastic ion currents and eventually to macroscopic fluctuations, here in the voltage of an excitable membrane. We emphasize that we now consider the characteristics of the current through an entire ion channel (\Ka or \Na), in contrast to the subunit activity addressed in the previous section.

The classical Hodgkin-Huxley model describes the dynamics of the membrane potential $V$ and the generation of an action potential in a neuron by means of a passive leak current, a voltage-dependent \Ka current, and a voltage-dependent \Na current \cite{Izh07, GerKis14}. In a stochastic formulation the latter two currents can be expressed by:
\begin{equation}
\begin{aligned}
    I_\text{K} &= g_\text{K} n^{(1)}(t)n^{(2)}(t)n^{(3)}(t)n^{(4)}(t)(V - E_\text{K}) \\
    I_\text{Na} &= g_\text{Na} m^{(1)}(t)m^{(2)}(t)m^{(3)}(t)h^{(1)}(t)  (V - E_\text{Na})
    \label{eq:HH_ion_currents}
\end{aligned}
\end{equation}
where $g_\text{K}$ and $g_\text{Na}$ are the maximal conductances and $E_\text{K}$ and $E_\text{Na}$ are the reversal potentials. In our case, the variables $n^{(i)}$, $m^{(j)}$, and $h^{(k)}$ are Markovian dichotomous processes that capture the state of the subunits in a single \Ka or \Na channel \footnote{The state is described as activated or deactivated for $n$ and $m$, and inactivated or deinactivated for $h$ \cite{DayAbb01}}. Accordingly to \eq{eq:HH_ion_currents}, the \Ka channel consists of four activation gates of type $n$, whereas the \Na channel consists of three activation gates of type $m$ and one inactivation gate of type $h$. Only when all subunits are open, the channel is open. 

In the original Hodgkin-Huxley model, the gating variables are deterministic quantities bounded between zero and one and governed by the differential equation:
\begin{align}
    \tau_x(V) \dot{x} = \alpha_x(V)(1-x) - \beta_x(V)x, 
    \label{eq:HH_gating_variables}
\end{align}
with $x=n,m,h$. In this case, $n$, $m$, and $h$ represent the fraction of open subunits in a large ensemble. However, eq.\ (\ref{eq:HH_gating_variables}) can also be interpreted as $x(t)$ describing the probability of a single subunit taking a certain state in a two-state system with voltage-dependent transition rates $\alpha_x(V)$ and $\beta_x(V)$ (similar to \fig{fig:1-diagram_dichotomous}). 
This insight allows to formulate stochastic variants of the Hodgkin-Huxley model consistent with the deterministic model in the thermodynamic limit, where the state of each subunit is represented by a Markovian dichotomous noise with a mean governed by \eq{eq:HH_gating_variables} \cite{FoxLu94, Fox97}. In this formulation, the gating variables correspond to the fluctuating fraction of open subunits in a finite ensemble. 

\begin{figure}
    \centering
    \includegraphics[scale=0.9]{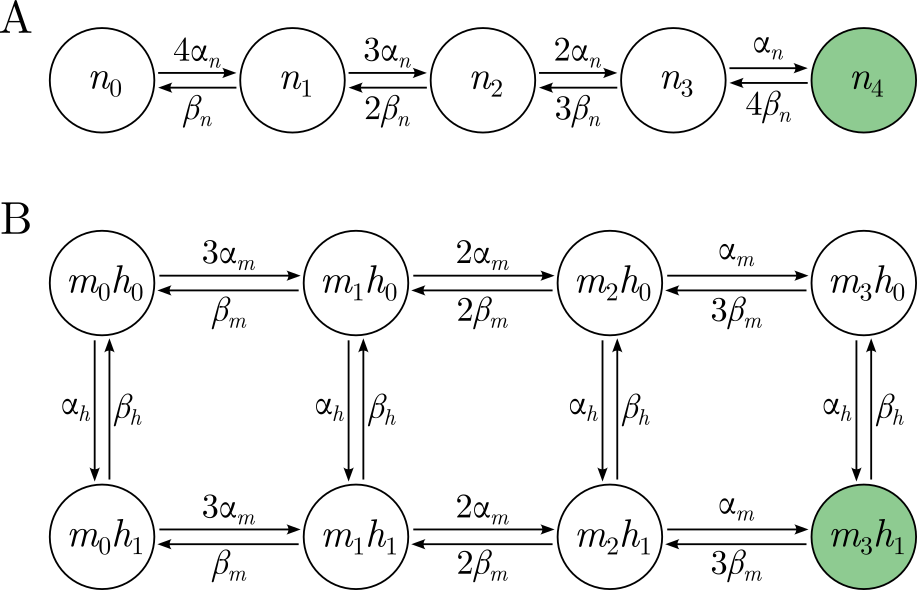}
    \caption{\textbf{State diagram of a \Ka channel and a \Na channel \cite{Koc99}.} Panel A shows the five-state model of a stochastic \Ka channel. The five states represent the number of activated $n$-type subunits (0 to 4) of the \Ka channel. The transition rates are given above/below the arrows.  All four subunits must be activated ($n_4$) for the \Ka channel to open. Panel B shows the eight-state model of a stochastic \Na channel. The eight states represent the number of activated $m$-type (0 to 3) and deinactivated $h$-type (0 to 1) subunits of the \Na channel. All three subunits of type $m$ must be activated and the subunit of type $h$ must be deinactivated ($m_3h_1$) for the \Na channel to open.}
    \label{fig:5-diagram-potassium-sodium-channels}
\end{figure}
\begin{table}
	\caption{\textbf{Simulation parameters for \Ka and \Na channels in the Hodgkin-Huxley model \cite{DayAbb01}}}
	\centering
    \begin{tabular}{ l c }
			\hline
			Parameter & Value \\
			\hline
            $\alpha_n$ / $\text{s}^{-1}$ & $0.01(V+55) / [1 - \exp(-0.1(V+55))]$ \\
			$\alpha_m$ / $\text{s}^{-1}$ & $0.1(V+40) / [1-\exp(-0.1(V+40))]$ \\
			$\alpha_h$ / $\text{s}^{-1}$ & $0.07 \exp(-0.05(V+65))$ \\
            $\beta_n$ / $\text{s}^{-1}$ & $0.125\exp(-0.0125(V+65))$ \\
            $\beta_m$ / $\text{s}^{-1}$ & $4 \exp(-0.0556(V+65))$ \\
            $\beta_h$ / $\text{s}^{-1}$ & $ 1 / [1 + \exp(-0.1(V+35))]$ \\
            $g_\text{K}$ / $\text{mS}$ & 36 \\
            $g_\text{Na}$ / $\text{mS}$ & 120 \\
            $E_\text{K}$ / $\text{mV}$ & -77 \\
            $E_\text{Na}$ / $\text{mV}$ &  50 \\
            \hline
		\end{tabular}
	\label{tab:parameters_potassium_sodum_channels}
\end{table}
In the following we calculate the noise intensity of the ion current through a single \Ka channel or a single \Na channel. We have already emphasized that the kinetics of a single subunit ($n^{(i)}(t)$, $m^{(j)}(t)$, and $h^{(k)}(t)$) can be described by a Markovian dichotomous noise with a transition rate matrix similar to the one used in \se{sec:dichotomous_noise}. One could be tempted to think that the noise intensity of the product of a number of independent random processes can be easily found from the intensities of the single processes. However, we are not aware of a simple relation between the former and the latter. 

\begin{figure}[t]
    \centering
    \includegraphics[width=\columnwidth]{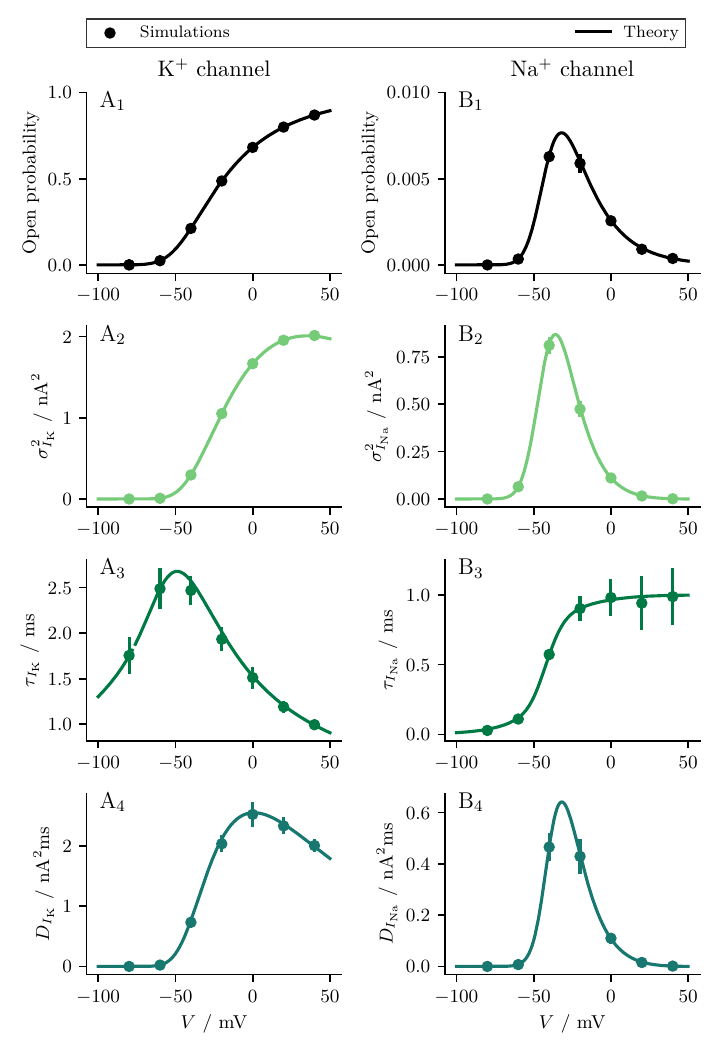}
    \caption{\textbf{Statistical measures of a stochastic \Ka and \Na channel.} Panels A and B show the open probability, variance $\sigma^2$, correlation time $\tau$, and noise intensity $D$ as a function of the membrane potential $V$ for the stochastic \Ka and \Na channels governed by the scheme illustrated in \fig{fig:5-diagram-potassium-sodium-channels} A and B, respectively. Vertical lines indicate the standard error calculated from ten simulations. The variance and noise intensity are calculated according to \eq{eq:mean_and_variance} and \eq{eq:noise-intensity_matrix}, respectively. The correlation time is determined as the ratio $\tau = D/\sigma^2$. Parameters are according to \tab{tab:parameters_potassium_sodum_channels}.}
    \label{fig:6-statistics_ka_na_channel}
\end{figure}
To calculate the noise intensity for the product, we need to formulate the transition rate matrix for the random processes $x(t) = n^{(1)}(t)n^{(2)}(t)n^{(3)}(t)n^{(4)}(t)$ for the \Ka channel or $x(t) = m^{(1)}(t)m^{(2)}(t)m^{(3)}(t)h^{(1)}(t)$ for the \Na channel. Here, we follow the formulation of \cite{Koc99} and use transition rate matrices corresponding to the state diagrams in \fig{fig:5-diagram-potassium-sodium-channels}. 
The five possible states for the \Ka channel (\fig{fig:5-diagram-potassium-sodium-channels}A) result from the fact that the subunits are identical and independent. 
Therefore, it is sufficient to describe the number of subunits in the activated state. In this formulation, the transition rate from the state $n_3$ (three activated $n$-type subunits) to $n_4$ is $\alpha_n$, the rate at which the last deactivated gate is activated, and the transition from $n_3$ to $n_2$ is $3\beta_n$, the rate at which one out of three activated gates deactivates (see \cite{Koc99, DayAbb01}). The entire \Ka channel is considered open when all gates are activated, i.e.\ the value of the state $n_4$ is $1$ and the value of every other state is $0$.
Similarly, a reduced state diagram can be formulated for the \Na channel (\fig{fig:5-diagram-potassium-sodium-channels}B). In this case, the number of activated $m$-type subunits and number of inactivated $h$-type subunit must be distinguished, resulting in eight different states. The \Ka channel is considered open, when all three $m$-type subunits are activated and the $h$-type subunits deinactivated, i.e.\ the value of the state $m_3 h_1$ is $1$ and the value of every other state is $0$.

In \fig{fig:6-statistics_ka_na_channel} we compare simulation results and theoretical predictions of the open probability, variance, correlation time, and noise intensity of the stochastic \Ka and \Na currents according to \eq{eq:HH_ion_currents} where the gating variables are described by the Markov schemes illustrated in \fig{fig:5-diagram-potassium-sodium-channels}A and \ref{fig:5-diagram-potassium-sodium-channels}B, respectively. In both cases, the numerically calculations agree with the theoretical predictions, demonstrating that the method is applicable to the stochastic Hodgkin-Huxley model. However, we note that our method relies on the assumption of a clamped voltage. 

Regarding the interpretation of the obtained curves, we first note that the two upper panels agree with the deterministic open probability of the classical Hodgkin-Huxley model: A monotonically increasing function for the \Ka channel (\fig{fig:6-statistics_ka_na_channel}A$_1$) and a non-monotonic function for the \Na channel (\fig{fig:6-statistics_ka_na_channel}B$_1$) due to the interplay between activation and inactivation. The latter maximum implies maxima in the variance (\fig{fig:6-statistics_ka_na_channel}B$_2$) and the noise intensity (\fig{fig:6-statistics_ka_na_channel}B$_4$). We note that there are also maxima in the characteristics of the \Ka channel at different voltage values (\fig{fig:6-statistics_ka_na_channel}A$_2$-A$_4$). The maximum of the variance is plausible because the open probability reaches zero and one in the limit of extreme voltage values. This maximum then also entails a maximum of the noise intensity. 

\section{Summary and discussion}
\label{sec:discussion}
In this paper, we have developed a general framework to characterize a noise process that is described by a finite Markov chain, i.e.\ by a Master equation with a finite number of states.
More specifically, we demonstrated that the calculation of the noise intensity and correlation time of the process is only slightly more complicated than the computation of the steady state and its mean and variance. We illustrated our general result by application to three cases: (i) the dichotomous noise (for which all characteristics are, of course, well known); (ii) a stochastic calcium channel subunit as it is used in the De Young-Keizer model; (iii) sodium and potassium currents as used in the Hodgkin-Huxley equation of action potential generation. In all these cases, our comparison to stochastic simulations of the underlying discrete dynamics agreed well with the analytical predictions of our formulas over a wide range of tested parameters.

The computation of noise intensity and correlation time has particular importance in the context of the so-called diffusion approximation. In several situations of interest the discrete fluctuations described by the Master equation can be well approximated by a white Gaussian noise. This stochastic process is fully characterized by its mean value and its noise intensity, for which we derived a simple expression above. 
Once this approximation has been made, the apparatus of nonlinear diffusion processes, in particular, the Fokker-Planck equation for the evolution of the probability density, can be used. In order to learn whether this approximation is really justified for a specific system, it is crucial to know the correlation time of the noise and to test whether it is much shorter than all other time scales in the system - only if this is the case, we are permitted to neglect the temporal correlations of noise  entirely. This may also apply in the more complicated situation in which both the mean and intensity depend on the dynamical variable(s) of the driven system, i.e.\ when there is a feedback between the dynamical variable(s) and the noise statistics.

Let us revisit the dynamics of the calcium subunit, for which the correlation time was shown in \bi{4-statistics_de-young-keizer}B as a function of the (clamped) calcium concentration.
The maximum correlation time is below 3 seconds in this case. If we now take into account that calcium is \emph{not} clamped but in fact obeys a dynamics on the time scale of tens of seconds to multiple minutes \cite{BouMar08, ThuSmi11}, we may justify to approximate the stochastic activity of the single subunit by a Gaussian white noise with a calcium-dependent mean value and a calcium-dependent noise intensity. This is true when the \Ca concentration is below some spiking threshold, and it does not include the dynamics that is responsible for the spike shape. For another calcium channel (cluster) model, this has been carried out in a integrate-and-fire type model of intracellular calcium excitability and thoroughly tested by us and a collaborator \cite{RamFal23,RamFal23b}. 

For the gating variables of the Hodgkin-Huxley model a similar argument may be possible. A naive version of the approximation is not justified to describe the generation of the action potential. It is exactly the nonlinear interplay between the voltage and gating dynamics that gives the action potential its characteristic shape, i.e.\ the upstroke of the spike caused by the positive feedback of sodium-channel opening upon an initial (and sufficiently strong) depolarization and the downstroke due to the slower inactivation of sodium channels and the opening of potassium channels. The membrane time constant in the original Hodgkin-Huxley model (roughly, the time scale of the voltage dynamics) is of the order of 3ms \cite{HodHux52} and thus comparable to the correlation time of the potassium channel fluctuations (according to \fig{fig:6-statistics_ka_na_channel}b around 2.5ms for voltage values around the resting potential). Hence, in this case it is recommendable to abstain from a white-noise approximation. Indeed, different approximation schemes that are based on having a large number of channels, have been divised, see e.g.\ the classical studies by Fox et al. \cite{FoxLu94, Fox97}, who approximated the gating dynamics by chemical Langevin equations, and more recent contributions which use stochastic shielding to obtain numerically efficient descriptions of the inherent stochasticity \cite{PuTho21}. We note that our results are still useful, because in experiment the voltage \emph{can and is routinely} clamped to a prescribed value and currents through specific channels can be isolated (methods for this are for instance discussed in the textbook by Izhikevich \cite{Izh07}), and in this situation our formulas give exact results for the characteristics of the respective current fluctuations.  The same method can be applied to more complicated kinetic schemes of channel states, see e.g.\ the review on the many models of sodium channels \cite{Pat91}.

Of course, the two above cases are more involved in the sense that not always the dynamics of the Markov chain itself is affected by the variable it drives. When the output of the Markov chain acts as an external noise on a system, no white-noise approximation has to be made and our results then simply provide the most important noise characteristics of this stochastic process, making it comparable to simpler noise models (white or low-pass filtered Gaussian noise, white Poissonian noise, or colored dichotomous noise). 

\providecommand{\noopsort}[1]{}\providecommand{\singleletter}[1]{#1}%

\end{document}